\documentclass[iop,twocolappendix,revtex4]{emulateapj}
\usepackage{graphicx,natbib,enumitem}
\usepackage{amsmath,amssymb,mathrsfs}
\usepackage{txfonts,color}

\def\tab{Table~}

\def\part{Section~}

\def\planck{\textit{Planck}{}}
\def\xmm{\textit{XMM-Newton}{}}
\def\chandra{\textit{Chandra}{}}

\def\T{\mathrm{T}}
\def\P{\mathrm{P}}
\def\Y{\mathrm{Y}}

\def\wt{\mathscr{Wt}}
\def\spl{\mathscr{S}}
\def\rec{\mathcal{R}}
\def\gauss{\mathcal{N}}

\begin{document}

\title{Pressure profiles of distant galaxy clusters in the \textit{Planck} catalog}
\author{H. Bourdin\altaffilmark{1,2}, P. Mazzotta\altaffilmark{2,1}, A. Kozmanyan\altaffilmark{2}, C. Jones\altaffilmark{1}, A. Vikhlinin\altaffilmark{1}}
\journalinfo{The Astrophysical Journal, 72:83 (12pp), 2017 July 1}
\submitted{Received 2017 March 14; accepted 2017 May 21; published 2017 July 5}

\newcommand{\myemail}{herve.bourdin@cfa.harvard.edu}

\shorttitle{Pressure profiles of distant galaxy clusters}
\shortauthors{Bourdin et al.}

\affil{{$^1$}{Harvard Smithsonian Centre for Astrophysics, 60 Garden Street, 
Cambridge, MA 02138, USA; \myemail}}
\affil{{$^2$}{Dipartimento di Fisica, Universit\`a degli Studi di Roma `Tor
  Vergata', via della Ricerca Scientifica, 1, I-00133 Roma, Italy}}

\begin{abstract}

	Successive releases of \planck~data have demonstrated the strength of the Sunyaev--Zeldovich (SZ) effect in detecting hot baryons out to the galaxy cluster peripheries. To infer the hot gas pressure structure from nearby galaxy clusters to more distant objects, we developed a parametric method that models the spectral energy distribution and spatial anisotropies of both the Galactic thermal dust and the Cosmic Microwave Background, that are mixed-up with the cluster SZ and dust signals. Taking advantage of the best angular resolution of the High Frequency Instrument channels (5 arcmin) and using X-ray priors in the innermost cluster regions that are not resolved with \planck, this modelling allowed us to analyze a sample of 61 nearby members of the \planck~catalog of SZ sources ($0 < z < 0.5$, $\tilde{z} = 0.15$) using the full mission data, as well as to examine a distant sample of 23 clusters ($0.5 < z < 1$, $\tilde{z} = 0.56$) that have been recently followed-up with \xmm~and \chandra~observations. We find that (i) the average shape of the mass-scaled pressure profiles agrees with results obtained by the \planck~collaboration in the nearby cluster sample, and that (ii) no sign of evolution is discernible between averaged pressure profiles of the low- and high-redshift cluster samples. In line with theoretical predictions for these halo masses and redshift ranges, the dispersion of individual profiles relative to a self-similar shape stays well below 10 \% inside $r_{500}$ but increases in the cluster outskirts.	
	
\end{abstract}

\keywords{Galaxy: clusters: general --- Galaxies: clusters: intracluster medium}

\section{Introduction}

Clusters of galaxies trace the most massive matter inhomogeneities that collapsed across cosmic times. The baryonic matter in galaxy clusters predominates in the form a hot ionized atmosphere that has reached virial temperatures, detectable at the same time from its X-ray bremsstrahlung emission and via the inverse Compton scattering of the Cosmic Microwave Background radiation (the so-called Sunyaev--Zeldovich effect, hereafter SZ). Because cluster atmospheres are thought to lie close to hydrostatic equilibrium, thermal gas pressure is the thermodynamical quantity that best relates their internal structure to the cluster masses. X-ray observations of a representative sample of the local cluster population have shown that this relation is particularly tight at intermediate cluster-centric radii~~\citep[$r_{2500} \le r \le r_{500}$\footnote{$r_{\Delta}$ is the radius of a ball with a density that equals $\Delta$ times the critical density of the Universe.},][hereafter A10]{Arnaud_10}, where the gas pressure structure exhibits a remarkable mass-scale invariance.  Reflecting the scale invariance of the underlying dark matter halos, the self-similarity of gas pressure profiles also motivates the use of empirical scaling relations that connect the total cluster masses to SZ and X-ray proxies of the averaged gas pressure, such as the integrated Compton parameter and its X-ray analog, $\mathrm{Y}_{\mathrm{x}}$ \citep{Kravtsov_06}.

The $\Lambda$-CDM cosmological scenario predicts that clusters grow via the continuous accretion of less-massive, not-yet-virialized structures of the cosmic web.  The hydrostatic equilibrium of cluster atmospheres is thus an approximation that must, at least, be corrected by perturbative terms that reflect the pressure support of anisotropic gas motions. Numerical $N$-body simulations show that the non-thermal pressure fraction related to these motions increases with cluster-centric radii \citep{Lau_09,Nelson_14}, and that its sensitivity to the mass-accretion-rate tends to break the self-similarity of the gas pressure profile in the cluster outskirts \citep[$r \ge r_{500}$,][]{Lau_15}. As a result, simulated cluster samples usually exhibit an increase of the thermal pressure scatter in the cluster outskirts (see e.g.  \citealt{McCarthy_14} and \citealt{Kravtsov_12} for a review of earlier works), at times coming with a steepening of the average profile shapes at the highest cluster redshifts \citep[][]{Battaglia_12,Lau_15}. From the observational side, a mild redshift steepening of averaged gas pressure profiles might be seen in stacked \chandra~observations of clusters detected in the SZ by the South Pole Telescope \citep{McDonald_14}. 

In contrast to X-ray surface brightness that quadratically depends on the gas density and that is affected by the source redshift dimming, the SZ Compton parameter is just proportional to the integrated gas pressure along the line of sight and does not depend on cluster redshifts. Therefore, the thermal SZ signal is a direct, weakly biased tracer of the pressure structure, that can easily explore the outer regions of individual clusters, regardless of their distance. After the completion of its nominal mission, the \planck~collaboration extracted the SZ signal of 62 nearby clusters detected at high significance in the \planck~catalog \citep{Planck_2013_V_pressure}. The radially averaged SZ signal in this sample extended up to $3 \times r_{500}$, with a shape consistent with X-ray-derived pressure profile below $r_{500}$, but slightly exceeding the theoretical predictions from cosmological simulations of cluster formation in the outskirts. Most clusters in this study were individually mapped beyond $r_{500}$, which demonstrated the capabilities of \planck~to map the Compton parameter out to the cluster outskirts.

\begin{table*}[!t]
\caption{Distant high SZ flux, X-ray observed subsample of the \planck~cluster catalogue. Cluster masses have been estimated from the SZ flux, assuming a flux-radius scaling relation calibrated using X-ray observations of nearby clusters \citep{Planck_2015_2ndszcat}.}
\begin{center}
\begin{tabular}{lrrrrrr}
\tableline\tableline
Target name  & Redshift       & Matched-filter & $\mathrm{r}_{500}$ &             $\mathrm{M}_{500}$ & \xmm & \chandra \\
             &                &            SNR &              (kpc) & ($10^{14} \mathrm{M}_{\odot}$) &      &          \\
\tableline
PSZ2 G265.10-59.50 & 0.50 &  6.7 & 1055 &   5.74& \checkmark &\\
PSZ2 G044.77-51.30 & 0.50 &  8.3 & 1239 &   9.32& \checkmark &\\
PSZ2 G211.21+38.66 & 0.50 &  5.6 & 1127 &   7.03& \checkmark &\\
PSZ2 G212.50-61.38 & 0.50 &  5.4 & 1125 &   7.00& \checkmark &\\
PSZ2 G110.28-87.48 & 0.52 &  6.6 & 1119 &   7.01& \checkmark &\\
PSZ2 G201.50-27.31 & 0.54 &  7.1 & 1220 &   9.27& \checkmark &\\
PSZ2 G094.56+51.03 & 0.54 &  7.8 & 1091 &   6.65& \checkmark &\\
PSZ2 G004.45-19.55 & 0.54 &  9.1 & 1305 &  11.38& \checkmark &\\
PSZ2 G228.16+75.20 & 0.55 & 11.4 & 1288 &  11.01&& \checkmark \\
PSZ2 G111.61-45.71 & 0.55 &  9.7 & 1230 &   9.58& \checkmark & \checkmark\\
PSZ2 G180.25+21.03 & 0.55 & 12.8 & 1357 &  12.89& \checkmark &\\
PSZ2 G183.90+42.99 & 0.56 &  5.8 & 1079 &   6.59& \checkmark &\\
PSZ2 G155.27-68.42 & 0.57 &  8.0 & 1210 &   9.35& \checkmark &\\
PSZ2 G239.93-39.97 & 0.58 &  6.6 & 1080 &   6.75& \checkmark &\\
PSZ2 G144.83+25.11 & 0.58 &  7.1 & 1140 &   7.99& \checkmark &\\
PSZ2 G260.63-28.94 & 0.60 &  7.8 & 1069 &   6.72& \checkmark & \checkmark\\
PSZ2 G073.31+67.52 & 0.61 &  6.4 & 1060 &   6.63& \checkmark &\\
PSZ2 G070.89+49.26 & 0.61 &  5.7 & 1071 &   6.84& \checkmark &\\
PSZ2 G045.87+57.70 & 0.61 &  5.4 & 1125 &   7.94& \checkmark &\\
PSZ2 G099.86+58.45 & 0.63 &  7.8 & 1071 &   7.00& \checkmark &\\
PSZ2 G219.89-34.39 & 0.70 &  6.3 & 1112 &   8.53& \checkmark &\\
PSZ2 G297.97-67.74 & 0.87 & 13.1 & 1129 &  10.94&& \checkmark \\
PSZ2 G266.54-27.31 & 0.97 &  7.8 & 1050 &   9.91&& \checkmark \\
\tableline
\tableline
\end{tabular}
\end{center}
\end{table*}

In order to investigate the evolution of pressure profile shapes, the present work aims to extend the \planck~collaboration study to higher cluster redshifts. We introduce a spatially variable modeling of the Galactic foreground and Cosmic Microwave Background (CMB) anisotropies, which enables us to exploit the 10- to 5-arcminutes angular resolution provided by the High Frequency Instrument (HFI) bolometers in the frequency range covered by the thermal SZ signal. Taking advantage of the statistical gain provided by the full 30-months mission data set and using X-ray priors in the innermost cluster regions that are not resolved with \planck, we extracted radial profiles of the thermal SZ signal toward the outskirts of the nearby ($0<z<0.5$) cluster sample presented by the \planck~collaboration and within a more distant ($0.5<z<1$) subsample of the \planck~catalogue. After describing these subsamples in Sect. 2, we detail our analysis of \xmm, \chandra~and \planck~data in Sect. 3. Our results are presented in Sect. 4 and their perspectives are discussed in Sect. 5. Hereafter, $m_p$ is the proton mass, $\mathrm{G}$ is the gravitational constant, $f_b$ is the mean baryonic fraction of the Universe, and $\mu$ and $\mu_e$ are the mean molecular weight and the mean molecular weight per free electron of the intra-cluster plasma. Moreover, intracluster distances are computed as angular diameter distances, assuming a $\Lambda$-CDM cosmology with $\mathrm{H}_\mathrm{0} =70~\mathrm{km}~\mathrm{s}^{-1}~\mathrm{Mpc}^{-1}$, $\Omega_{\mathrm{M}} = 0.3$, $\Omega_{\Lambda} = 0.7$. Unless otherwise noted, confidence intervals and envelopes encompass a 68 \% probability.

\section{Cluster samples \label{sect:cluster_samples}}

\begin{figure}[!t]
   \begin{center}
  \resizebox{\hsize}{!}{\includegraphics{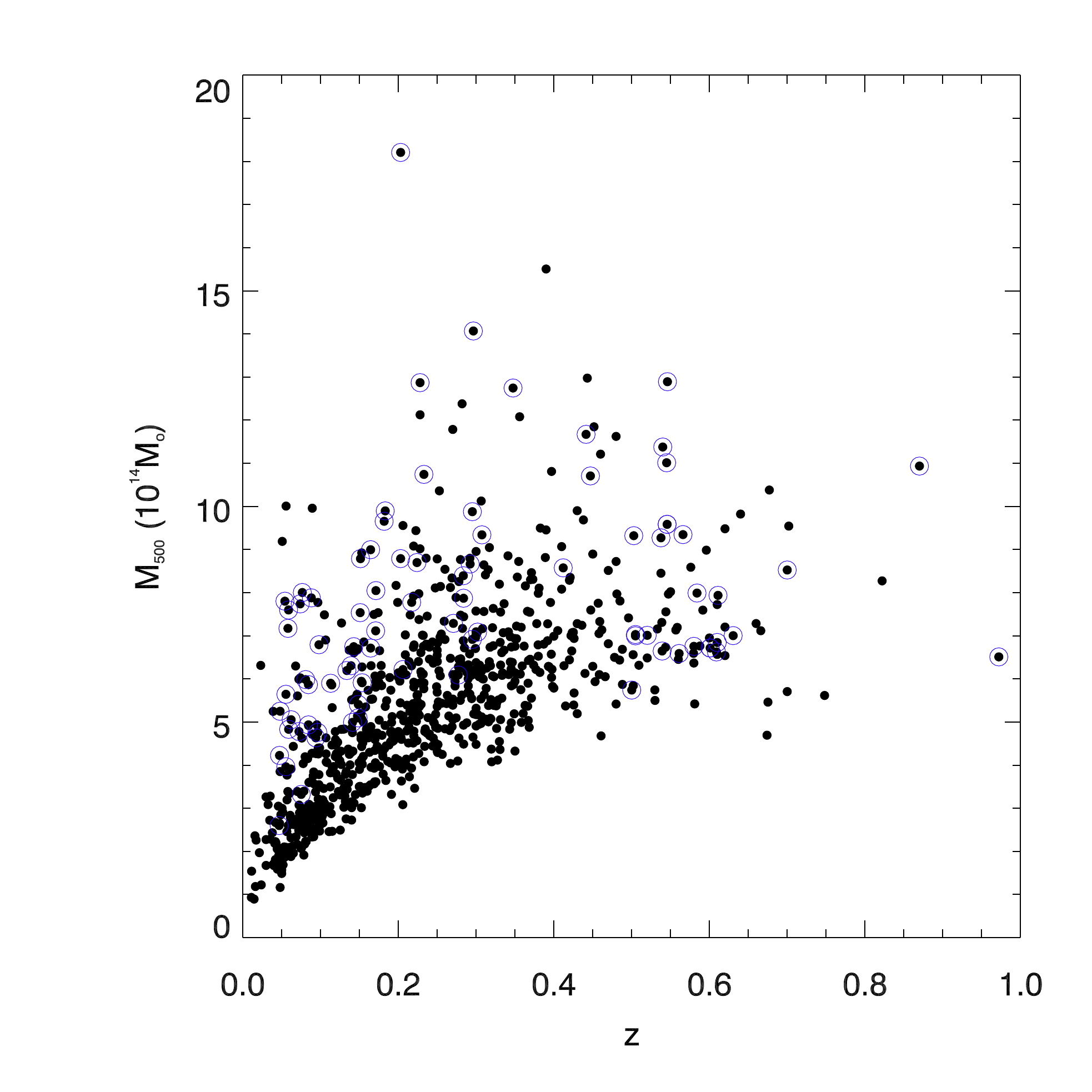}}
  \caption{Mass-redshift distribution of our galaxy cluster sample.  \textit{Black points:} Members of the \planck~catalogue of SZ sources with a matched-filter SNR that exceeds 5. \textit{Blue circles:} targets of the present work. \label{Fig:szcat_mz}}
   \end{center}
\end{figure}

	After completion of its full mission, the \planck~collaboration released an all-sky catalog of SZ sources that contains 1653 detections, including 1203 confirmed clusters with identified counterparts \citep{Planck_2015_2ndszcat}. This catalog relies on cluster detections using a Multifrequency Matched Filter (MMF) approach, with their associated Signal-to-Noise Ratio (SNR). For this work we use two subsamples of the \planck~catalogue that will be detailed below: the low redshift sample and the high redshift sample. Cluster masses have been estimated from the SZ flux measured within $5 \times r_{500}$, assuming a flux-radius scaling relation calibrated using X-ray observations of nearby clusters \citep{Planck_2011_XI_szscaling}. As shown in Fig. \ref{Fig:szcat_mz}, both of these subsamples span a nearly uniform mass distribution across redshifts.

\subsubsection*{Low redshift sample}
The low redshift sample coincides with the sample used by the \planck~collaboration to calibrate SZ-scaling relations with X-ray observables and cluster masses. Members of this sample were selected for both their high SNR in the nominal 14-months data set, and the high quality of available X-ray follow-ups. A full description of their properties is provided in \planck~Collaboration (2011). One member of this sample, ZwCl 1215+0400, has been removed from the present analysis as a result of its missing detection in the second Planck catalog of SZ sources. Following the MMF definition, these nearby clusters are now detected at a SNR higher than 7 in the final 30-month \planck~data set. They cover a redshift range of $0.047<z<0.447$ and a mass range of $2.6<\mathrm{M}_{500}/{10^{14} \mathrm{M_{\odot}}<18.2}$ , with median values of $z$=0.15 and $\mathrm{M}_{500} = 7.17 \times 10^{14} \mathrm{M_{\odot}}$, respectively.

\subsubsection*{High redshift sample}
The high redshift sample is composed of 23 clusters detected above $z=0.5$ with a SNR higher than 5, for which reliable X-ray data is available in the \xmm~or \chandra~archives. \tab 1 lists the redshifts and characteristic radii of these distant clusters. The sample spans a redshift range of $0.5<z<1$ and a mass range of $5.7<\mathrm{M}_{500}/{10^{14} \mathrm{M_{\odot}}<12.9}$, with median values of $z$=0.56 and $\mathrm{M}_{500} = 7.9 \times 10^{14} \mathrm{M_{\odot}}$, respectively. 

\begin{figure*}[!ht]
   \begin{center}
 \resizebox{\hsize}{!}{\includegraphics{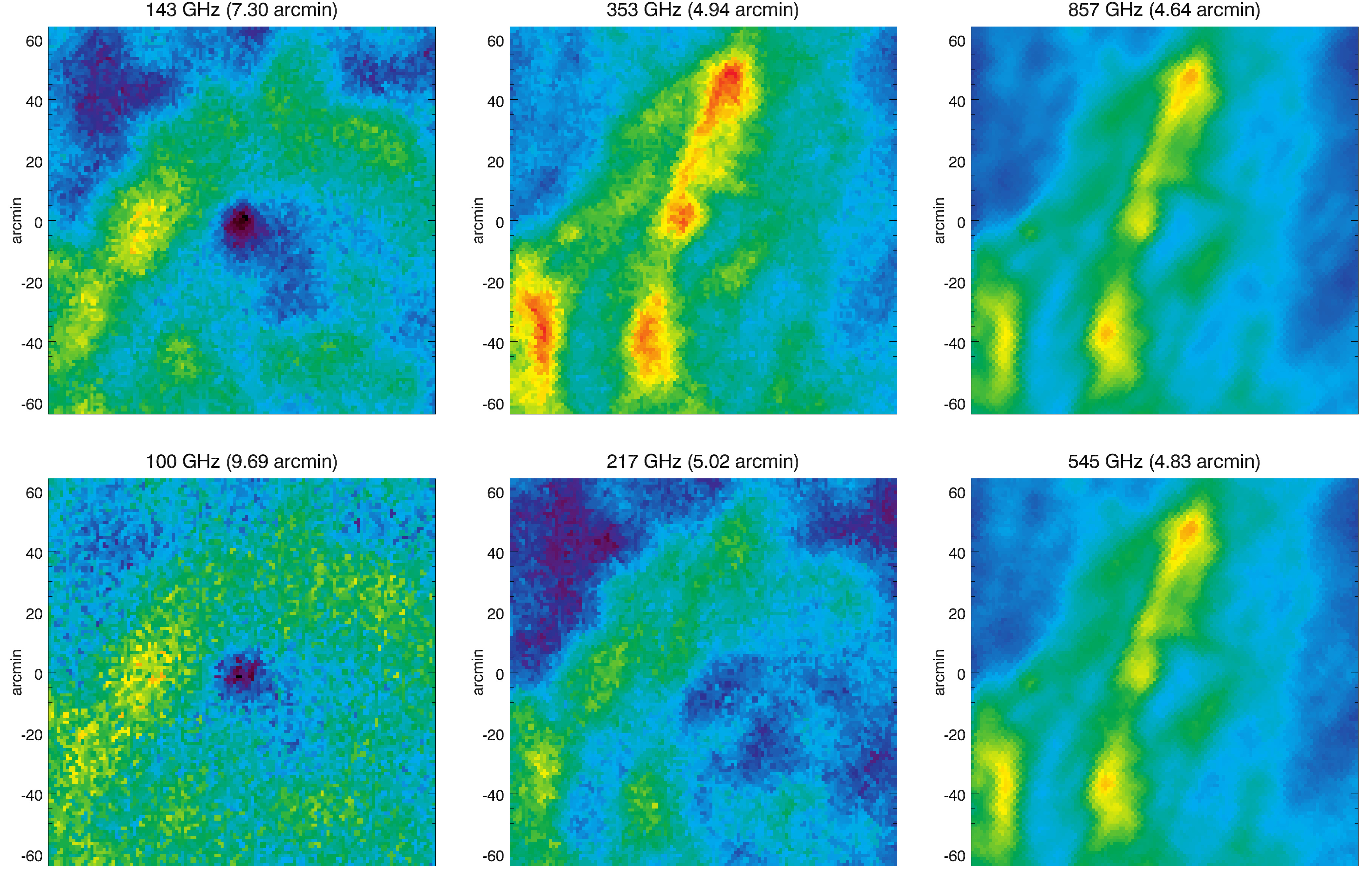}}
  \caption{\planck-HFI frequency maps in the neighborhood of the galaxy cluster A2163. Numbers in brackets stand for the angular resolution. The cluster signal is primarily superimposed onto temperature anisotropies of the CMB at low frequencies ($\nu \le 143$ GHz), and spatial variations of the Galactic thermal dust (GTD) emissivity and high frequencies ($\nu \ge 353$ GHz ). Both CMB and GTD anisotropies contribute to the pattern visible at 217 GHZ.\label{Fig:a2163_fmaps}}
   \end{center}
\end{figure*}

\begin{figure*}[!ht]
   \begin{center}
  \resizebox{\hsize}{!}{\includegraphics{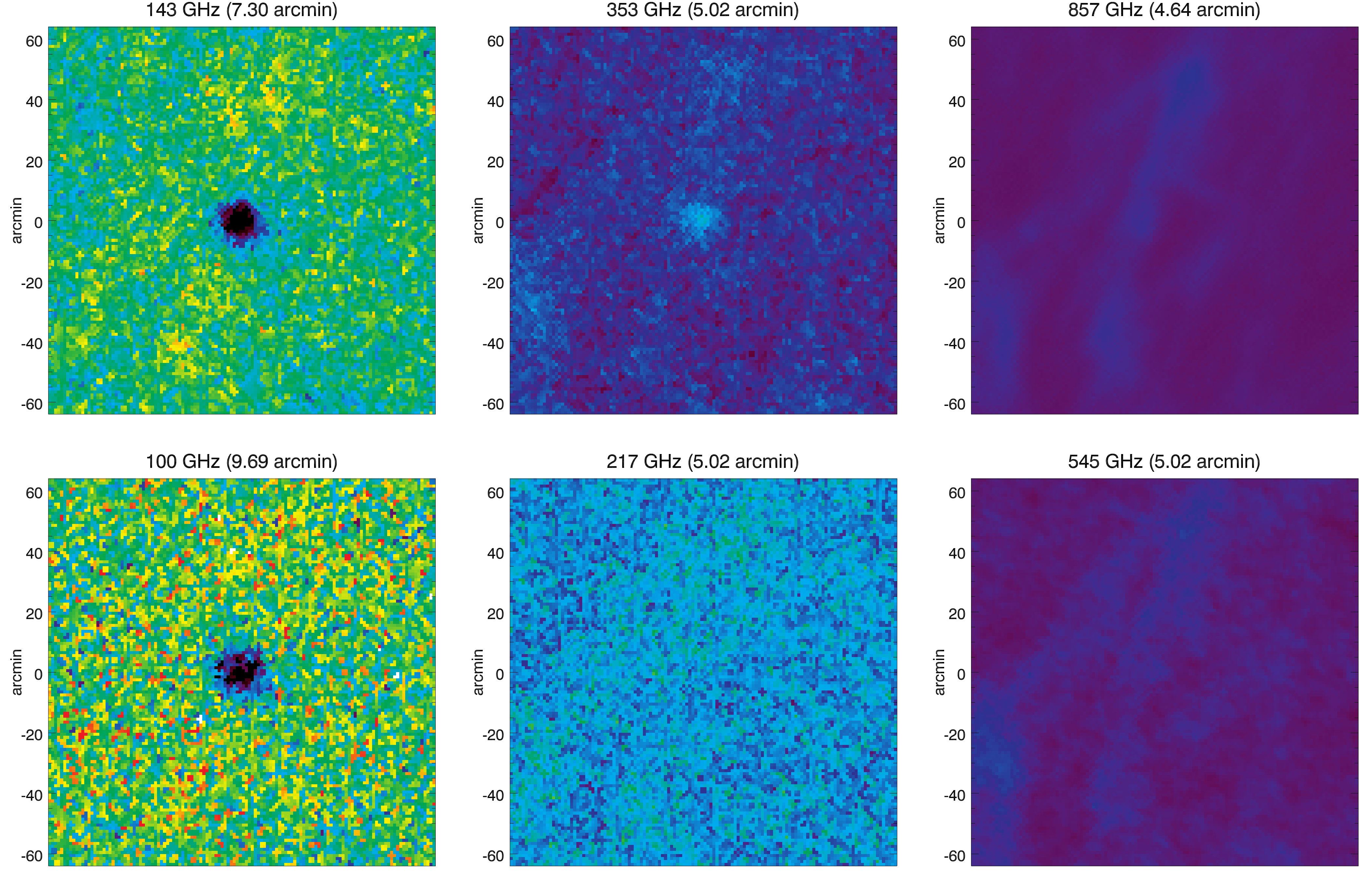}}
  \caption{Millimetric signal detected toward the galaxy cluster Abell 2163. Maps result from a high-pass filtering of the \planck-HFI frequency maps (see Fig. \ref{Fig:a2163_fmaps}) and a subtraction of Cosmic Microwave Background and Galactic thermal dust anisotropies  (see details in Section \ref{subsect:millimetric}). This subtraction predominantly reveals the thermal SZ signal --a decrement at 100 and 143 GHz, an increment at 353 GHZ-- and residual contributions from the cluster thermal dust emissivity.\label{Fig:a2163_szsignal}}
  \end{center}
\end{figure*}

\begin{figure*}[!ht]
   \begin{center}
\resizebox{\hsize}{!}{\includegraphics{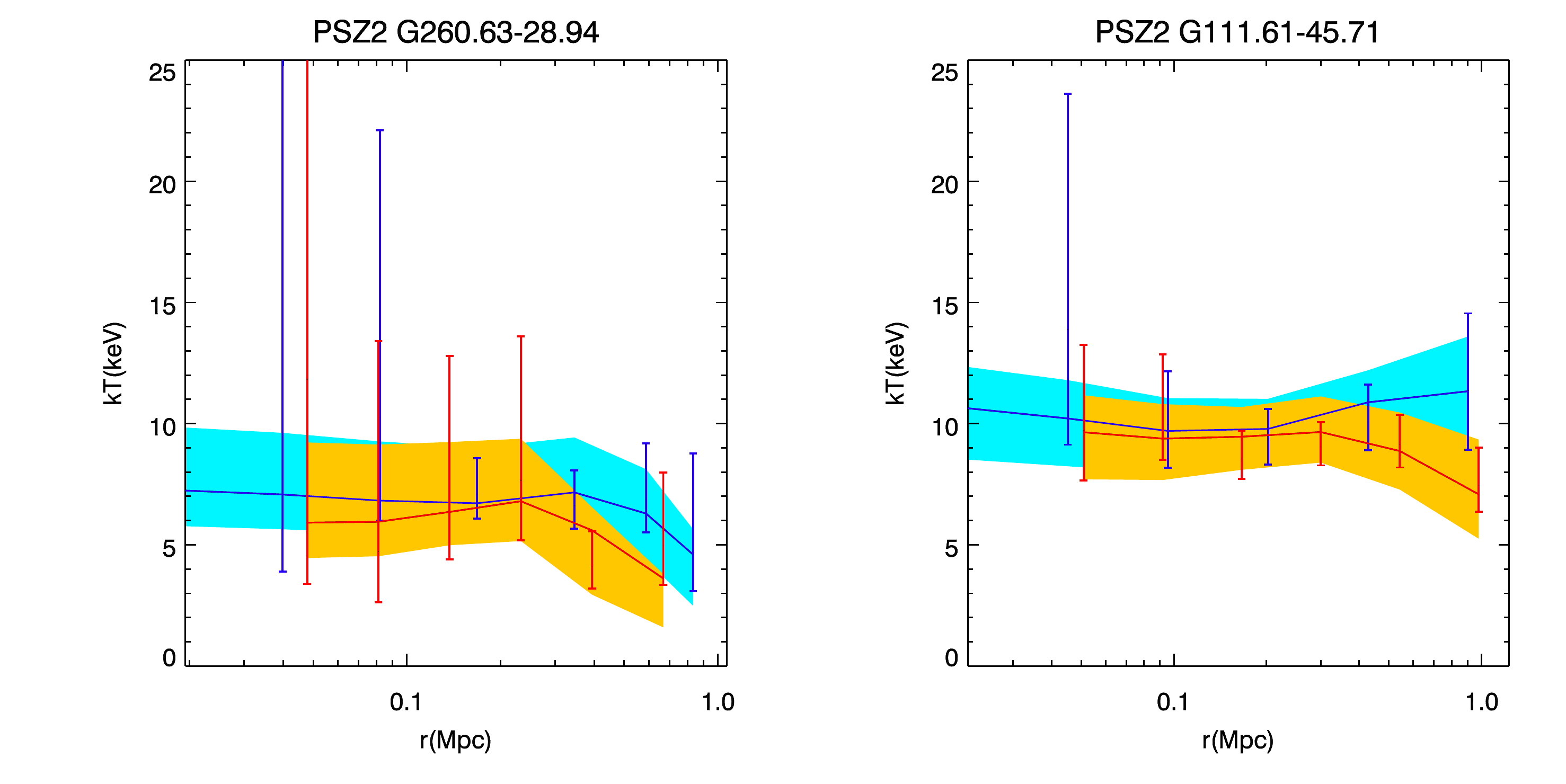}}
  \resizebox{\hsize}{!}{\includegraphics{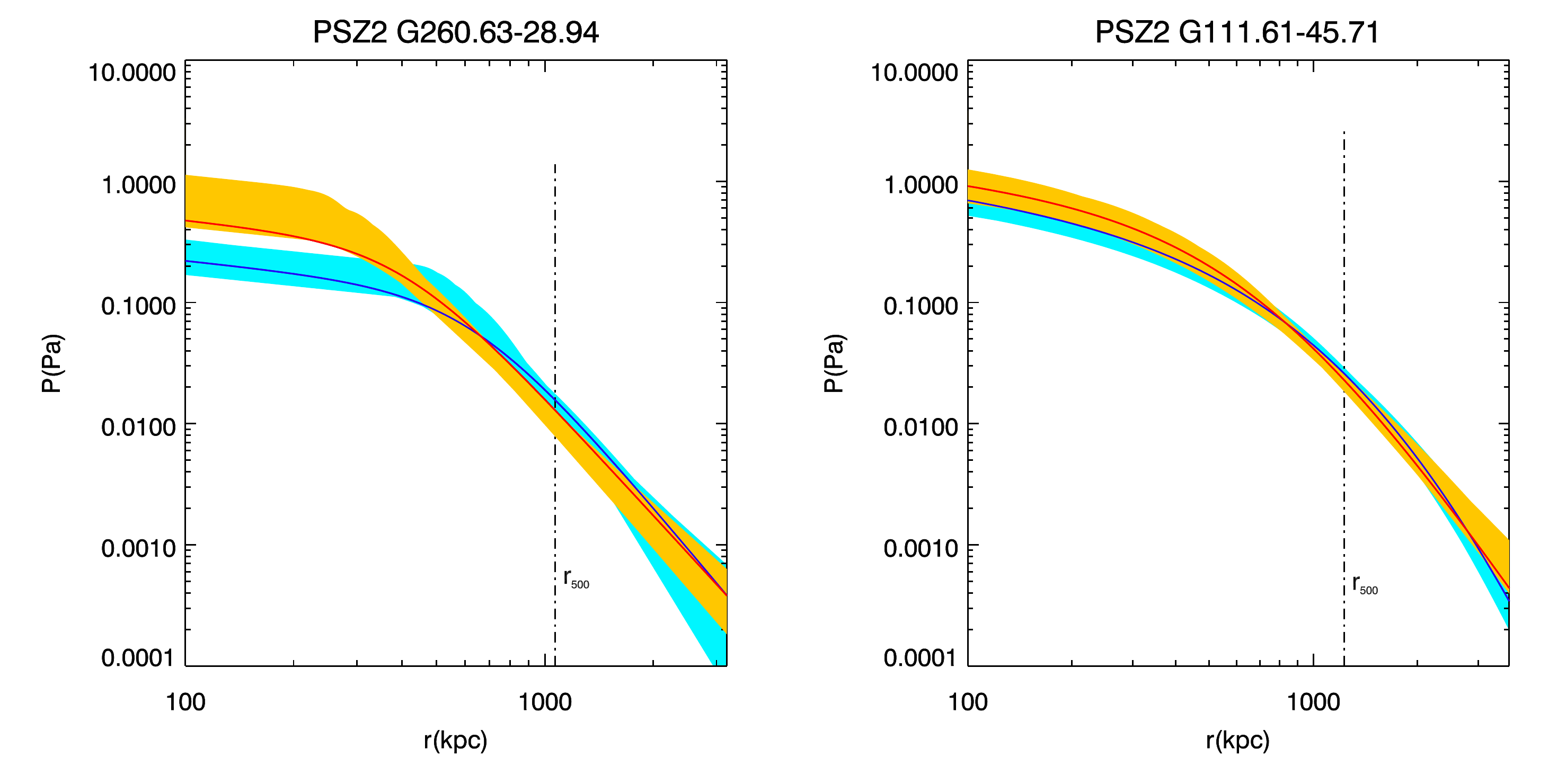}}
  \caption{Joint spectroscopic temperature (top) and gas pressure (bottom) fits to the millimetric and X-ray data pointing toward PSZ2G111.61-45.71 and PSZ2G60.63-28.94. \textit{Red-orange:} \xmm~and \planck~data. \textit{Blue:} \chandra~and \planck~data.\label{Fig:joint_tpfit}}
   \end{center}
\end{figure*}

\section{Data analysis\label{sect:data_analysis}}

\subsection{X-ray data:\label{sect:xray_data}}

The X-ray emission in our cluster samples has been probed using \xmm~and \chandra~observations performed with the European Photon Imaging Camera (EPIC) and Advanced CCD Imaging Spectrometer, respectively. We extracted radial profiles of the soft [0.5, 2]~keV X-ray surface brightness, $\Sigma_x$, and the spectroscopic temperature, $\T_x$, that were used to invert spherically symmetric templates of the gas emission measure, $[n_p n_e](r)$, and temperature, $\T(r)$. These templates were modeled using analytical forms first proposed in \citet{Vikhlinin_06}:

\begin{eqnarray}
  [n_p n_e](r) &=& 
  \frac{(r/r_c)^{-\alpha^\prime}}{[1+(r/r_c)^2]^{3\beta_1 - \alpha^\prime/2}}
  \frac{n_{0}^2}{[1+(r/r_s)^3]^{\epsilon/3}} \nonumber \\ &&+ \frac
  {n_{02}^2}{[1+(r/r_{c2})^2]^{3\beta_2}},  \label{rho3d_equ} \\
  \T(r) &=& \T_0 (x+\T_{min}/\T_0) / (x+1) \frac{(r/r_t)^{-a}}{(1+(r/r_t)^b)^{c/b}},
\end{eqnarray}

where $x = (r/r_{cool})^{a_{cool}}$. These analytical forms were subsequently integrated along the line of sight and fitted to the X-ray observables, $\Sigma_x$ and $\T_x$,  assuming the temperature-weighting scheme proposed in \citet{Mazzotta_04} in order to mimic the spectroscopic response of a single temperature fit

\begin{eqnarray}
 {\Sigma_x}(r) &=& \frac{1}{4 \pi (1+z^4)}  \int [n_p n_e](r) \Lambda(T,Z) dl \\
 {\T_x(r)} &=& \frac {\int w \T(r) dl}{\int w dl}, 
 \label{tsl_equ}
\end{eqnarray}

with $w={n_e^2}/{T^{3/4}}$. In this work, $\Sigma_x$ and $\T_x$ were centered onto the X-ray image peak and gathered photon events in radial bins where wavelet-detected point sources have been excluded. Secondary cluster substructures were also masked out in a few complex systems. As detailed in \citet{Bourdin_08}, such measurements rely on spatially variable, energy-dependent effective areas and detector responses. They invoke a background noise model that includes false detections due to the interaction of high-energy particles with the detectors and astrophysical components that are combined with the cluster signal, such as a cosmic X-ray background and a (two-temperature) Galactic foreground. These components, the spatial and spectral shape of which are detailed in \citet{Bourdin_13} and \citet{Bartalucci_14}, were spectrally fitted and normalized in an external annulus where no cluster emission is expected ($r>1.4\times r_{500}$), apart for a few very nearby systems that cover the field of view. For these nearby systems, the background components were jointly fitted to the cluster emissivity in the outskirts.

The inversion of $[n_p n_e](r)$ and $\T(r)$ assumes the redshifted, Galactic-hydrogen-absorbed Spectral Energy Distribution (SED) of the hot gas combines a bremsstrahlung continuum with metal emission lines tabulated in the Astrophysical Plasma Emission Code \citep[APEC,][]{Smith_01}. By adopting the solar composition of metal abundances tabulated by \citet{Grevesse_98} and a constant normalization of 0.3, this is achieved using a parametric bootstrap where $ {\Sigma_x}$ is convolved with the mirror Point-Spread Function (PSF) in the case of \xmm, and fitted to several surface brightness and temperature realizations. This approach typically yields temperature and emission measure profiles with relative uncertainties of about $10\%$ and $1\%$, respectively. As shown in \citet{Martino_14}, emission measure profiles, $[n_pn_e](r)$, turn out to be robust to X-ray telescope cross-calibration issues, suggesting that \chandra~and \xmm~effective areas are consistent with each other in the soft band, and that the \xmm~PSF smearing is well known and invertible. Adopting the metal composition detailed above and a particle mean weight of $\mu=0.596$ allows us to infer an average ratio of proton and electron densities of $n_p/n_e$ = 0.852. We use this value to infer an electronic density distribution, $n_e(r)$, from the parametric form of $[n_pn_e](r)$ given in equation (\ref{rho3d_equ}).

\subsection{Millimetric and submillimetric data:\label{subsect:millimetric}}

	The galaxy cluster signal has been extracted using \planck~HFI data from the full 30-months mission. All-sky HFI frequency maps have been reprojected toward smaller, eight square-degree maps, centered on each cluster. In these maps, the SZ and dust cluster signal is combined with Galactic foreground and CMB anisotropies on cluster scales, while cosmic infrared and SZ backgrounds contribute on larger scales together with instrumental offsets. To isolate the millimetric cluster signal from the other components, we applied the following procedure:

\begin{enumerate}[leftmargin=*,align=left]
\item We reduce the angular resolution of HFI maps in the range 217--857 GHz, to a common value of 5 arcmin.  We further convolve each frequency map, ${I}_{HFI}(\nu)$, with a third-order B-spline\footnote{see \citealt{Curry_47} for definition.} kernel, $\spl$, with a typical width of 1 degree and 15 arcmin for the nearby and distant cluster samples, respectively. This procedure yields a smoothed image that we subtract from the raw image. The resulting maps, $\tilde{I}_{HFI}(\nu)$, are filtered from anisotropies with scales that exceed the cluster size, including the astrophysical backgrounds and instrumental offsets:
\begin{equation}
\tilde{I}_{HFI}(\nu) =  \left[ \delta - \spl \right] \ast {I}_{HFI}(\nu),
\label{hfi_highpass}
\end{equation}

where $\delta$ is the Dirac distribution.
\item To build a spatial template of the Galactic thermal dust anisotropies, $I_{GTD}(857)$, we denoise the 857 GHz frequency map via a wavelet coefficient thresholding. We compute an isotropic undecimated wavelet transform \citep{Starck_07} of $\tilde{I}_{HFI}(\nu)$, that decomposes into B3-spline wavelet coefficients and a reconstruction that is obtained by a simple co-addition of the wavelet bands. Coefficient thresholds are preliminarily set to a 99.7 \% confidence level from noise simulations that match the variance and power-spectrum inferred from differences of the half-ring data sets. As expected for a spatially correlated noise \citep{Johnstone_97,Mallat_99}, they depend on both wavelet scale and position.
Denoting $\rec_{\wt}$ as the image reconstruction operator associated with thresholded wavelet coefficients, $\bar{\wt} \left[\tilde{I}_{HFI}(857)\right]$, we get:
\begin{equation}
{I}_{GTD}(857) =  \rec_{\wt} \bar{\wt}~\left[\tilde{I}_{HFI}(857)\right]
\end{equation}

\item As proposed by \citet{Meisner_15}, we model the SED of thermal dust anisotropies as a two temperature graybody that contains enough free parameters to fit the HFI data at all frequencies:
\begin{eqnarray}
	s_{GTD}(\nu) &=& \left[ f_1 q_1/q_2 \left(\frac{\nu}{\nu_o}\right)^{\beta_{d,1}} B_\nu(T_1) \right. \nonumber \\ 
				&& \left. + ~(1-f_1) \left(\frac{\nu}{\nu_o}\right)^{\beta_{d,2}} B_\nu(T_2)  \right] \nonumber \\ 
	I_{GTD}(\nu) &=& \eta_{GTD} I_{GTD}(857) \frac{s_{GTD}(\nu)}{s_{GTD}(857)}
\label{galactic_dust_equ}
\end{eqnarray}
Parameters of this model are the two graybody temperatures, $T_1$ and $T_2$, the two spectral indexes, $\beta_{d,1}$ and $\beta_{d,2}$, and the cold component fraction, $f_1$. The cold component fraction $f_1$ and spectral index are fixed a priori from their average all-sky value, while the temperatures, $T_2$, and $T_1=f(T_2, q_1/q_2,\beta_{d,1},\beta_{d,2})$ have been mapped a priori from a joint fit to \planck~and IRAS all-sky maps.

\item By subtracting our thermal dust template from the 217 GHz frequency map, we infer a spatial template of the CMB anisotropies, $I_{CMB}(217)$. In the same way as for thermal dust, this template is denoised via a wavelet thresholding and extrapolated to all HFI frequencies, following the average SED of the CMB dipole:
\begin{eqnarray}
{I}_{CMB}(217) &=&  \rec_{\wt} \bar{\wt} \left[\tilde{I}_{HFI}(217) - I_{GTD}(217)\right] \nonumber \\  
{I}_{CMB}(\nu) &=&  \eta_{CMB} {I}_{CMB}(217) \frac{B_\nu(T_{CMB})}{B_{217}(T_{CMB})}
\end{eqnarray}

\item We refine the thermal dust SED in the neighborhood of each cluster. To do so, we jointly fit the CMB template, $I_{CMB}(\nu)$, and the thermal dust template, $I_{GTD}(\nu)$, to the data set extracted in the cluster-centric radii range [7, 12] $r_{500}$, where no cluster signal is expected. The free parameters of this fit are the cold component fraction, $f_1$, and spectral index, $\beta_{d,1}$, plus the overall template normalizations $\eta_{GTD}$ and $\eta_{CMB}$.

\item By means of convolutions with Gaussian beams of characteristic width, $FWHM(\nu) = \sqrt{{FWHM_\nu}^2-{FWHM_{217}}^2}$, we reduce the angular resolution of the dust and CMB templates down to their expected resolution, $FWHM_\nu$, in the frequency range 100 to 217 GHz. We subsequently subtract these templates from the HFI data set. The dust and CMB-subtracted data set is the cluster signal, $I_{cluster}(\nu)$, which combines the thermal SZ signal with a secondary contribution related to the difference between Galactic and the intra-cluster dust SEDs: 
\begin{equation}
I_{cluster}(\nu) = \tilde{I}_{HFI}(\nu) - I_{GTD}(\nu) - I_{CMB}(\nu)
\label{mmcluster_equ}
\end{equation}
 
 \end{enumerate}

In this modeling, the correction of each SED for the HFI spectral response includes a unit conversion factor between the HFI 100-353 GHz channels and the 545 and 857 GHz channels that are calibrated in units of CMB temperature and intensities of a power-law SED, respectively \citep[see details in][]{Planck_2013_VIII_hfimaps,Planck_2013_IX_hfiresp}. A color correction is further applied to adapt the Galactic dust SED to the power-law used to calibrate the high-energy channels. These corrections are calculated using the Unit conversion and Color Correction (UcCC) package provided with the current \planck~data release.

An example of millimetric signal extraction is provided in Fig. \ref{Fig:a2163_fmaps} and \ref{Fig:a2163_szsignal} for the hot, massive galaxy cluster A2163. In the frequency range [353--857] GHz, the HFI maps show a filamentary template of thermal dust, while the CMB anisotropies and the thermal SZ distortion are mostly visible at lower frequencies. After high-pass filtering of the maps and subtraction of the thermal dust and CMB templates, the SZ signal becomes clearly visible in the frequency range [100-353] GHz.

\begin{figure*}[!htb]
   \begin{center}
 \resizebox{\hsize}{!}{\includegraphics{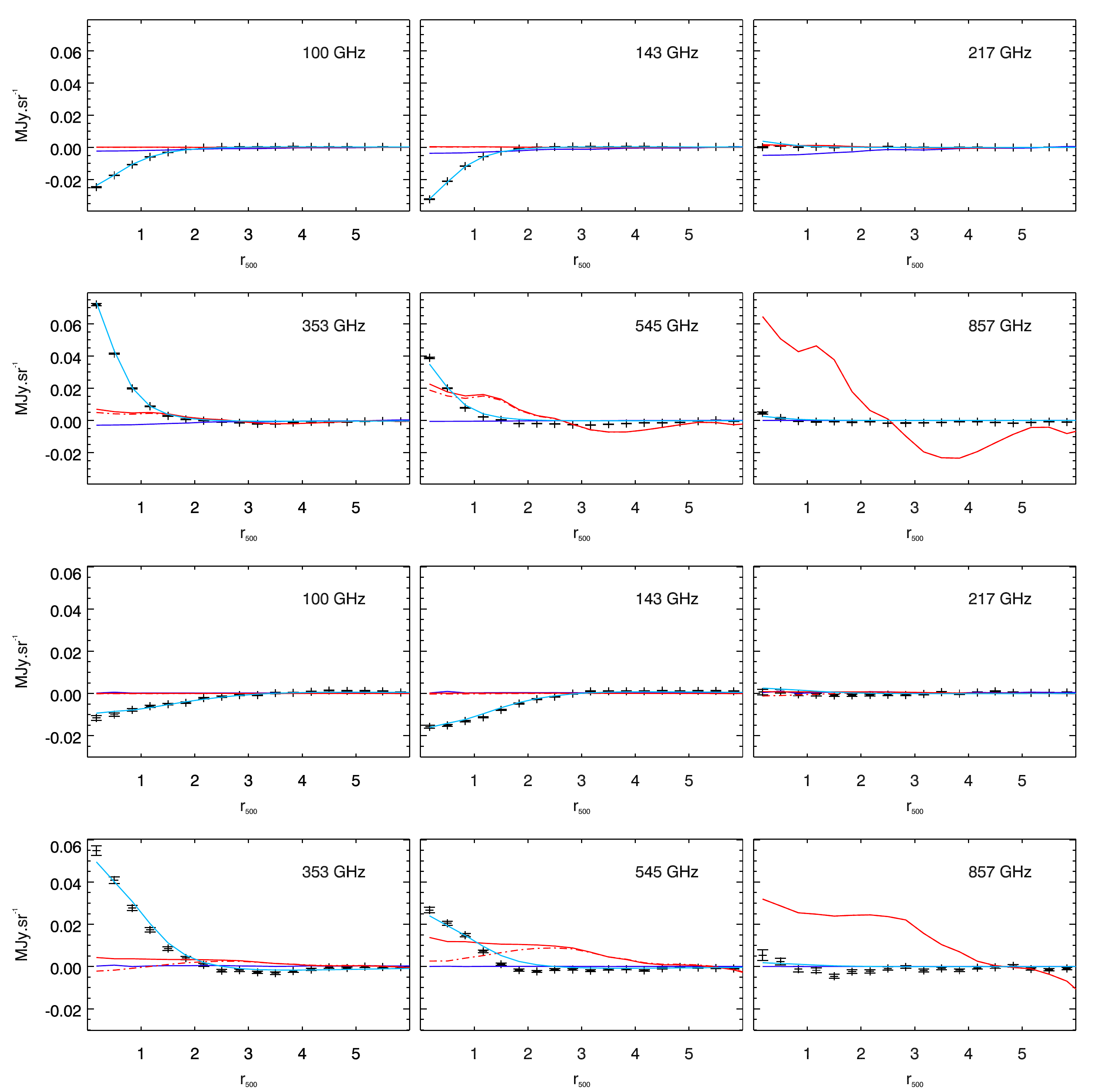}}
    \caption{Stacked radial profiles of the thermal SZ signal toward the nearby (top panels) and distant (bottom panels) cluster samples.  \textit{Dark blue line:} average CMB template. \textit{Red Continuous line:} Average Galactic thermal dust template. \textit{Red Dotted-Dashed line:}  Average correction of the thermal dust template for intra-cluster emissivities. \textit{Black points:} Stacked HFI data corrected for CMB and thermal dust anisotropies. \textit{Light blue curve:} Average model of the thermal SZ signal.\label{Fig:stacked_freqprofiles}}
   \end{center}
\end{figure*}

\begin{figure*}[!htb]
   \begin{center}
 \resizebox{\hsize}{!}{\includegraphics{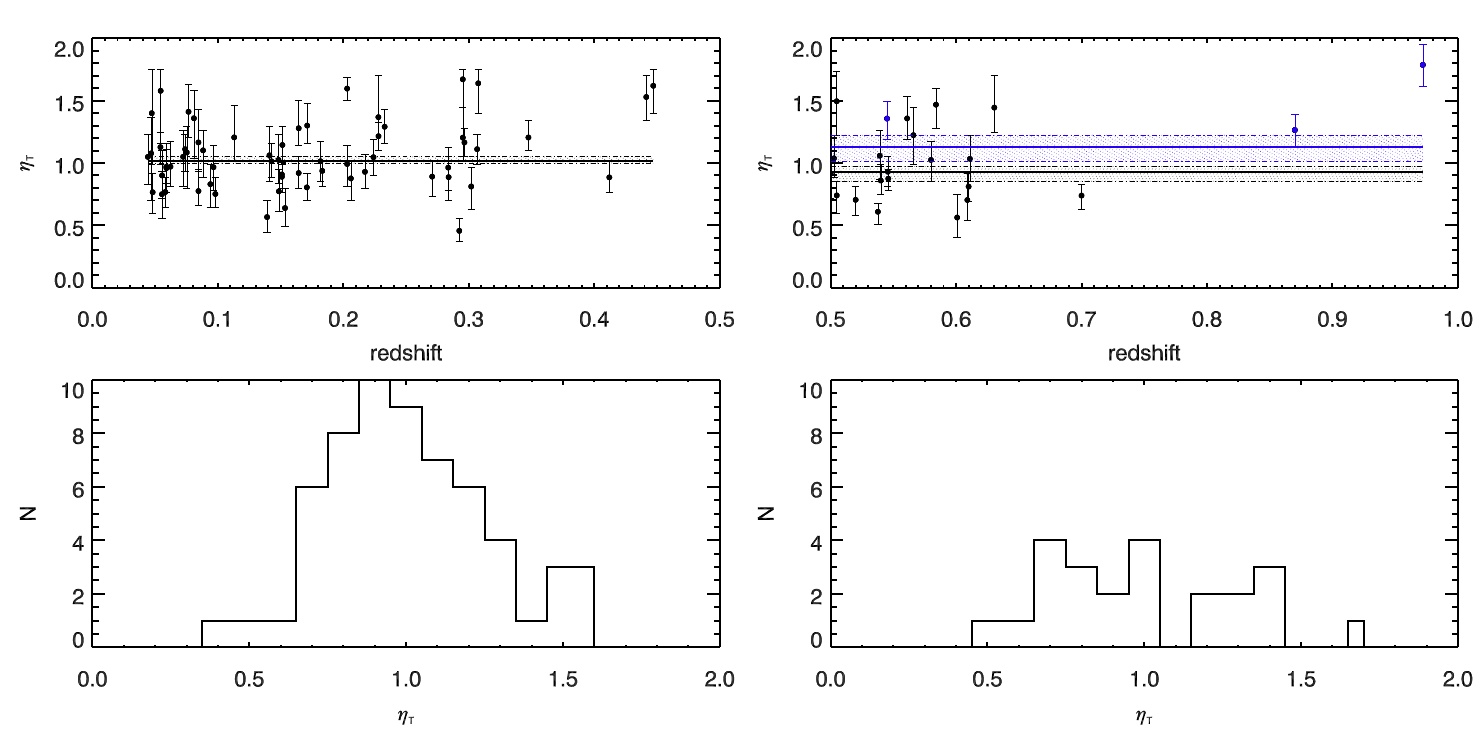}}
  \caption{Normalization of the spectroscopic temperatures measured in the nearby (left) and distant (right) cluster samples. \textit{Top:} Normalization values. Black and blue points correspond to \xmm~and \chandra~measurements, respectively. Horizontal lines depict temperature normalizations of the average profiles in each sample. \textit{Bottom:} Histogram of the normalization values. \label{fig:tx0_values}}
   \end{center}
\end{figure*}

\subsection{Joint X-Ray and SZ extraction of the cluster pressure profiles: \label{subsect:joint_xszpressure}}

\subsubsection*{Individual cluster profiles}
  
	In order to probe the gas pressure structure in both the inner and outer cluster regions, we perform the joint fit of a cluster pressure profile to the X-ray and millimeter signals.  For each cluster, we assume that the pressure structure follows a spherically symmetric distribution, $\P(r)$, first proposed by \citet[][hereafter N07]{Nagai_07}: 
	
\begin{eqnarray}
   	\P(r) = P_{0} \times \frac{\P_{500}}{x^{\gamma}(1+x^\alpha)^{(\beta-\gamma)\alpha}},
	\label{equ:nagai_07}
\end{eqnarray} 
   
   where $x = r/r_s$, $r_s=r_{500}/c_{500}$.  Given the millimetric and X-ray data sets, the inversion of $\P(r)$ is twofold:
   
\begin{enumerate}[leftmargin=*,align=left]

\item To fit the millimetric data, $\P(r)$ is used to build the thermal SZ signal expected for each HFI frequency, $I_{SZ}(\nu)$. Specifically, we integrate $\P(r)$ along the line of sight and multiply the ensuing map by a thermal SZ Kompaneets non-relativistic distortion of the CMB spectrum:
\begin{eqnarray}
	s_{SZ}(\nu) &=& \frac{h\nu}{kT} \left[ \frac{exp(h\nu/{kT})+1}{exp(h\nu/{kT})-1} - 4 \right], \nonumber \\
	I_{SZ}(\nu) &=&  s_{SZ}(\nu) ~ \frac{\sigma_T}{m_ec^2}    \int P(r) dl.
\end{eqnarray}

To take into account the cluster thermal dust emissivity, we further introduce a corrective term, $I_{CTD}(\nu)$, that will be detailed below. These components are convolved with HFI beams and high-pass filters that have been applied to the HFi frequency maps (Equ. \ref{hfi_highpass}):

\begin{equation}
\widehat{I_{cluster}(\nu)}  =  \gauss(FWHM_\nu) \ast \left[ \delta - \spl \right] \ast  \left[  I_{SZ}(\nu) +I_{CTD}(\nu) \right]
\label{equ:millimetric_template}
\end{equation}

The resulting cluster template, $\widehat{I_{cluster}(\nu)}$, is fitted to the millimetric cluster signal, as defined in Eq. (\ref{mmcluster_equ}).

\item To fit the X-ray data, $\P(r)$ is used to build a spectroscopic temperature template that derives from the ideal gas law:
\begin{equation}
k \T(r) = \eta_\T \times \P(r)/n_e(r),
\end{equation}
where $n_e(r)$ is determined a priori from the X-ray analysis (see Equ. \ref{rho3d_equ}) and $\eta_\T$ is a normalization constant. This template is integrated along the line if sight following Equ. (\ref{tsl_equ}), then fitted to a radial profile of projected X-ray temperatures extracted in the innermost cluster region (typically inside $r_x \sim r_{500}$).  
\end{enumerate}

In this modeling, any characteristic uncertainty around $n_e(r)$ is ignored as a result of its negligible amplitude with respect to spectroscopic temperature fluctuations. Moreover, introducing a free normalization, $\eta_\T$, to the spectroscopic temperatures aims at bypassing any bias in their X-ray estimates. Model-driven origins of such a bias include the presupposed values of the Hubble constant and the intracluster particle mean weight, or any average departure from the cluster spherical symmetry. Data-driven systematics  likely responsible for such a bias include cross-calibration mismatches between the X-ray instruments as reported for instance in \citet{Snowden_08,Martino_14}, and \citet{Schellenberger_15}.

\begin{table*}[!t]
\caption{Parameters of the average pressure profiles fitted to the stacked X-ray and SZ data sets in our cluster samples.}
\begin{center}
\begin{tabular}{ccccccccc}
\tableline\tableline
Cluster sample  & $\eta_{T,XMM}$  & $\eta_{T,Chandra}$  & ${P}_0$  & $c_{500}$  & $\gamma$ & $\alpha$ &  $\beta$   & $<f(M)>$\\
\tableline
$z<0.5$    & $1.01^{+0.02}_{-0.04}$   &  --      &  $5.25^{+0.23}_{-0.10}$    &  $1.18^{+0.02}_{-0.02}$    &  $0.31^{+0.00}_{-0.00}$    &  $1.27^{+0.01}_{-0.02}$    &  $5.41^{+0.06}_{-0.09}$    &  1.11\\
\tableline
$z>0.5$   & $0.92^{+0.07}_{-0.05}$   &  $1.13^{+0.12}_{-0.09}$      &  $6.23^{+0.51}_{-0.97}$    &  $1.11^{+0.00}_{-0.16}$    &  $0.31^{+0.00}_{-0.00}$    &  $1.12^{+0.02}_{-0.16}$    &  $5.36^{+0.38}_{-0.01}$    &  1.13\\
\tableline\tableline
\end{tabular}
\end{center}
\end{table*}

Analyses of stacked frequency maps toward clusters of the \planck~catalogue have shown that a thermal dust signal of cluster origin is detectable in the HFI data \citep{Planck_15_SZCIB,Planck_16_clusterdust}. This signal is expected to spatially overlap with the SZ signal and to follow an average SED, $s_{CTD}(\nu)$,  that is similar to the Galactic SED once corrected for the cluster redshifts. To take into account this perturbation of the SZ signal, we follow the Planck collaboration works and model $s_{CTD}(\nu)$ as a graybody characterised by spectral index, $\beta_d = 1.5$ and the intrinsic temperature, $T_d = 19.2 \mathrm{K}$. Its apparent temperature is corrected for each cluster redshift via $T_d^\prime=T_d/(1+z)$. Moreover, its spatial distribution is the projection of  a Navarro--Frenk--White \citep[][]{Navarro_97} dust density distribution along the line of sight, $\rho_{NFW}(r)$, with the concentration parameter $c_{500}=1$. To introduce $s_{CTD}(\nu)$ into the cluster template of Equ. (\ref{equ:millimetric_template}), we note that the cluster signal that derives from Equ. (\ref{galactic_dust_equ}) and (\ref{mmcluster_equ}) already subtract any thermal dust anisotropy at 857 GHz, and hence includes both Galactic and cluster thermal dust contributions at this frequency. It follows that the cluster template, $\widehat{I_{cluster}(\nu)}$, combines the thermal SZ signal with a linear combination of $s_{GTD}(\nu)$ and $s_{CTD}(\nu)$ that reaches zero at 857 GHz:

\begin{eqnarray}
s_{CTD}(\nu) &=& \nu^{\beta_d} B_\nu(T_d^\prime) \nonumber \\
I_{CTD}(\nu) &=& \left[ \frac{s_{CTD}(\nu)}{s_{CTD}(857)} - \frac{s_{GTD}(\nu)}{s_{GTD}(857)}\right]  \int \rho_{NFW}(r) dl \nonumber \\
\label{cluster_dust_equ}
\end{eqnarray}

The fit of $\P(r)$ and $\T(r)$ is performed by leaving $\P_0$, $\alpha$ and $\beta$ to vary, in addition to $\eta_\T$ and the amplitude of the cluster thermal dust contribution, $I_{CTD}(\nu)$. It is achieved via the joint minimization of millimetric and X-ray driven $\chi^2$, the parameter space of which is explored using Monte-Carlo Markov chains. Note that the millimetric driven $\chi^2$ includes covariance matrices that have been evaluated from instrumental noise simulations that preserve the power-spectrum measured, at each frequency, on differential maps combining both half-mission data sets. 
   
Fig. \ref{Fig:joint_tpfit} illustrates our pressure profile extraction for two galaxy clusters observed with both \chandra~and \xmm. The spectroscopic temperature measurements are slightly lower with \xmm~than with \chandra, with a decrement of about $10\%$ on radial average, that may reach $25\%$ in the outermost radial bins. Despite this discrepancy, the shape of the temperature profiles is consistent for most radial bins regardless of the X-ray telescope used, so that a joint fit of renormalized spectroscopic temperatures with \planck~data yields consistent pressure profiles.
   
\begin{figure*}[!tb]
   \begin{center}
  \resizebox{\hsize}{!}{\includegraphics{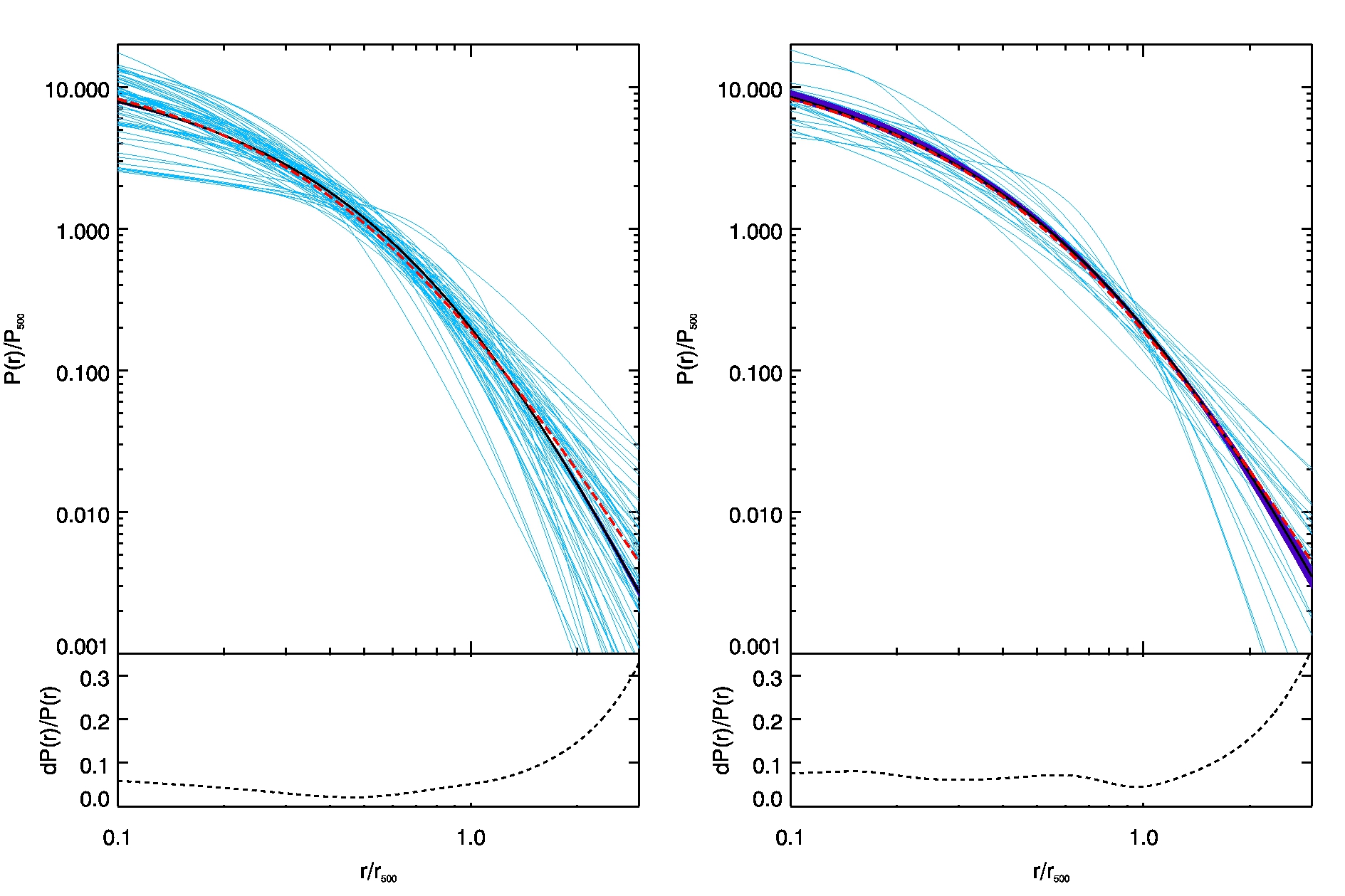}}
  \caption{Top panels: mass-scaled gas pressure profiles in the nearby (left) and distant (right) cluster samples. \textit{Light blue:} individual pressure profiles. \textit{Violet:} confidence envelope over an average pressure profile fitted to the stacked data set. \textit{Red dashed line:} \citet{Planck_2013_V_pressure} pressure template. Bottom panels: dispersion of individual profiles relative to the stacked profile values. \label{fig:massscaled_pressure}}
   \end{center}
\end{figure*}

\subsubsection*{Stacked mass-scaled profiles}

	For both the nearby and distant cluster sample defined in Sect. 2, we evaluate an average mass-scaled pressure profile via the fit of a unique self-similar function, $p(r) = \P(r) / \mathrm{P}_{500}$ to all X-ray and SZ data sets. Given a self-similar model of gravitational collapse (\citealt{Kaiser_86}), such a rescaling involves a renormalization of individual pressure templates to the characteristic gas pressure, $\mathrm{P}_{500} = n_{e,500} \mathrm{kT}_{500}$ (N07, A10), where $n_{e,500} = 500 f_b \rho_c(z) / (\mu_e \mathrm{m}_p$) and $k\T_{500} = \mu m_p G \mathrm{M}_{500} / (2 r_{500})$. Following N07 and A10, we conventionally adopt $f_b = 0.175$, $\mu  = 0.59$ and $\mu_e = 1.14$, which yields
\begin{equation}
\mathrm{P}_{500} = 1.65 \times 10^{-3} E(z)^{8/3} \times \left[ \frac{\mathrm{M}_{500}}{3 \times 10 ^{14} h_{70}^{-1} \mathrm{M}_{\odot}} \right]^{2/3} h_{70}^{2} ~\mathrm{keV}\mathrm{cm}^{-3}
\label{equ:P500}
\end{equation}
	where $E^2(z) = \Omega_M(1+z)^3 + \Omega_{\Lambda}$. The estimate of $p(r)$ is performed via the minimization of the sum of all individual $\chi^2$ values separating $\P(r)$ and $\T(r) = \eta_\T \times \P(r)/n_e(r)$ from the data, assuming a specific X-ray-derived density profile for each cluster, $n_e(r)$, and a specific temperature normalization $\eta_\T$ for each of the two X-ray instruments, \xmm~and \chandra. In this procedure, HFI data are corrected for the CMB and thermal dust templates that have been jointly fitted with the pressure profile of each individual cluster. Moreover,  $\chi^2$ values take into account spatial correlations of the instrumental noise in the same way as for individual clusters.

\section{Results\label{sect:Results}}

\subsection{Intracluster Component Separation, Hot Gas SZ Signal and Dust Emissivity}

For each target of our cluster samples, the component separation approach detailed in Sect. \ref{subsect:millimetric} and \ref{subsect:joint_xszpressure} yields models of the CMB and thermal dust anisotropies, plus a residual cluster signal obtained from subtraction of these components to the high-pass filtered frequency maps. For both our low- and high-redshift samples, Fig. \ref{Fig:stacked_freqprofiles} exhibits stacked radial profiles of these components centred toward clusters. As expected, the thermal SZ signal (light blue curve) averaged from individual pressure profile fits matches the stacked HFI data (black points) at all frequencies. The difference between the continuous and dashed red lines depicts the  correction of thermal dust anisotropies, $I_{CTD}(\nu)~$(see Equ. \ref{cluster_dust_equ}), that has been jointly fit to this signal. This correction, mainly visible at 545 GHz, reaches the order of magnitude of the thermal SZ signal in the high redshift sample, but remains relatively weaker at low redshift. This result is consistent with a picture in which the intrinsic averaged SED of the cluster member galaxies is similar in shape to the SED of our Galaxy, but exhibits a detectable redshift in distant clusters.

\subsection{Hot Gas Temperature}

As detailed in Sect. \ref{subsect:joint_xszpressure}, the joint X-ray and SZ extraction of the cluster pressure profiles provides us with a renormalisation of X-ray spectroscopic measurements, via the $\eta_\T$ parameter value. Distributions and histograms of these values measured for each cluster in our cluster samples are shown in Fig. \ref{fig:tx0_values}. The median values of $\eta_\T$  are 1.01 and 1.02 for the low- and high-redshift samples, respectively, with standard deviations of 0.27 and 0.33. Stacked data sets provide us with unique averaged value for each sample and instrument, $\eta_\T = 1.02^{+0.02}_{-0.03}$ for the nearby cluster sample, $\eta_\T = 0.92^{+0.05}_{-0.07}$ as for the distant cluster sample as seen by \xmm~and \planck, $\eta_\T = 1.13^{+0.09}_{-0.12}$ for the distant cluster sample as seen by \chandra~and \planck. These values are statistically consistent with $\eta_\T \equiv 1$, a value that would be expected in the absence of any bias on the Hubble constant value, for an ideal intracluster gas showing spherically symmetric density and pressure distributions. Under these assumptions they also agree with the consistency observed in the low-redshift sample between the \xmm~and \planck~estimates of the integrated gas pressure in the innermost cluster regions, via the $\Y_{X,500}$ and $\Y_{500}$ proxies \citep{Planck_2011_XI_szscaling}. 

\begin{figure}[t]
   \begin{center}
	\resizebox{\hsize}{!}{\includegraphics{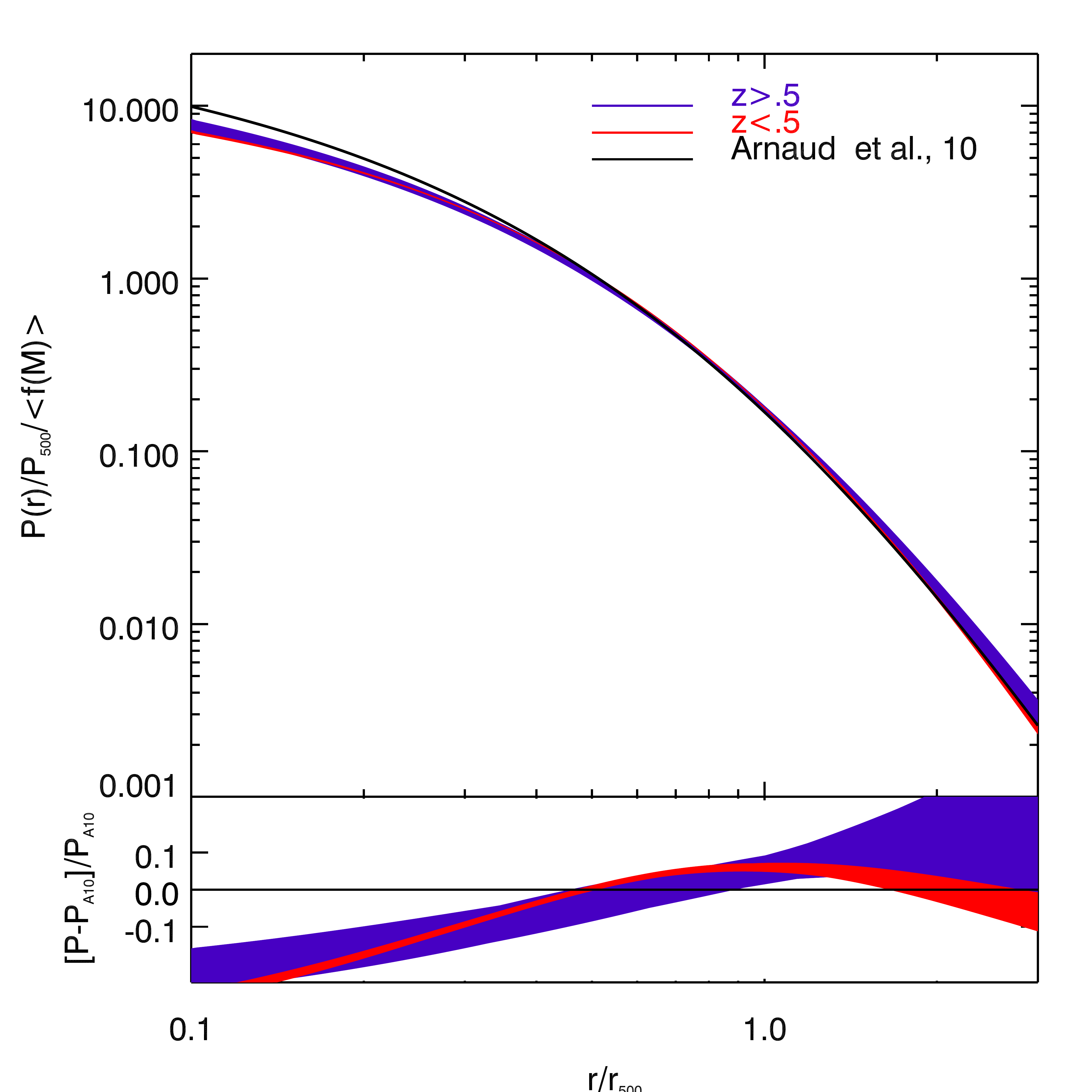}}
  \caption{\textit{Top panel}: stacked mass-scaled pressure profiles in the nearby (red) and distant (violet) cluster samples superimposed onto the analytical form proposed by \citet[][herafter, A10]{Arnaud_10}. \textit{Bottom panel}: relative difference between observed profiles and the analytical form of A10. \label{fig:massscaled_pressure_evol}}
   \end{center}
\end{figure}

\subsection{Mass-scaled pressure profiles}

\subsubsection*{Profile shape and dispersion}

	 Fig. \ref{fig:massscaled_pressure} exhibits the individual gas pressure profiles evaluated for each cluster in our samples. These profiles are re-scaled by their characteristic gas pressure, $\mathrm{P}_{500}$, and shown together with average scaled profiles and their confidence envelopes, the parameters of which are given in Table 2. The stacked profile in the nearby cluster sample is very close to the average profile obtained from the nominal 14-months data set by the \planck~collaboration, using a Modified Inter Linear Combination Algorithm \citep{Hurier_13} to derive Compton parameters maps. Specifically, the \planck~collaboration and the present estimate are fully consistent in the range $[0.1,2] \times r_{500}$, while the present estimate is marginally steeper outside $2 \times r_{500}$. The relative dispersion of individual profiles with respect to the stacked profile values, ${[\P(r) - \mathrm{\bar{P(r)}}]} / {\bar{P(r)}}$, is always lower than $8\%$ inside $r_{500}$, but increases in the cluster outskirts and reaches $15\%$ at $2 \times r_{500}$. 
		    
\begin{figure}[t]
   \begin{center}
  \resizebox{\hsize}{!}{\includegraphics{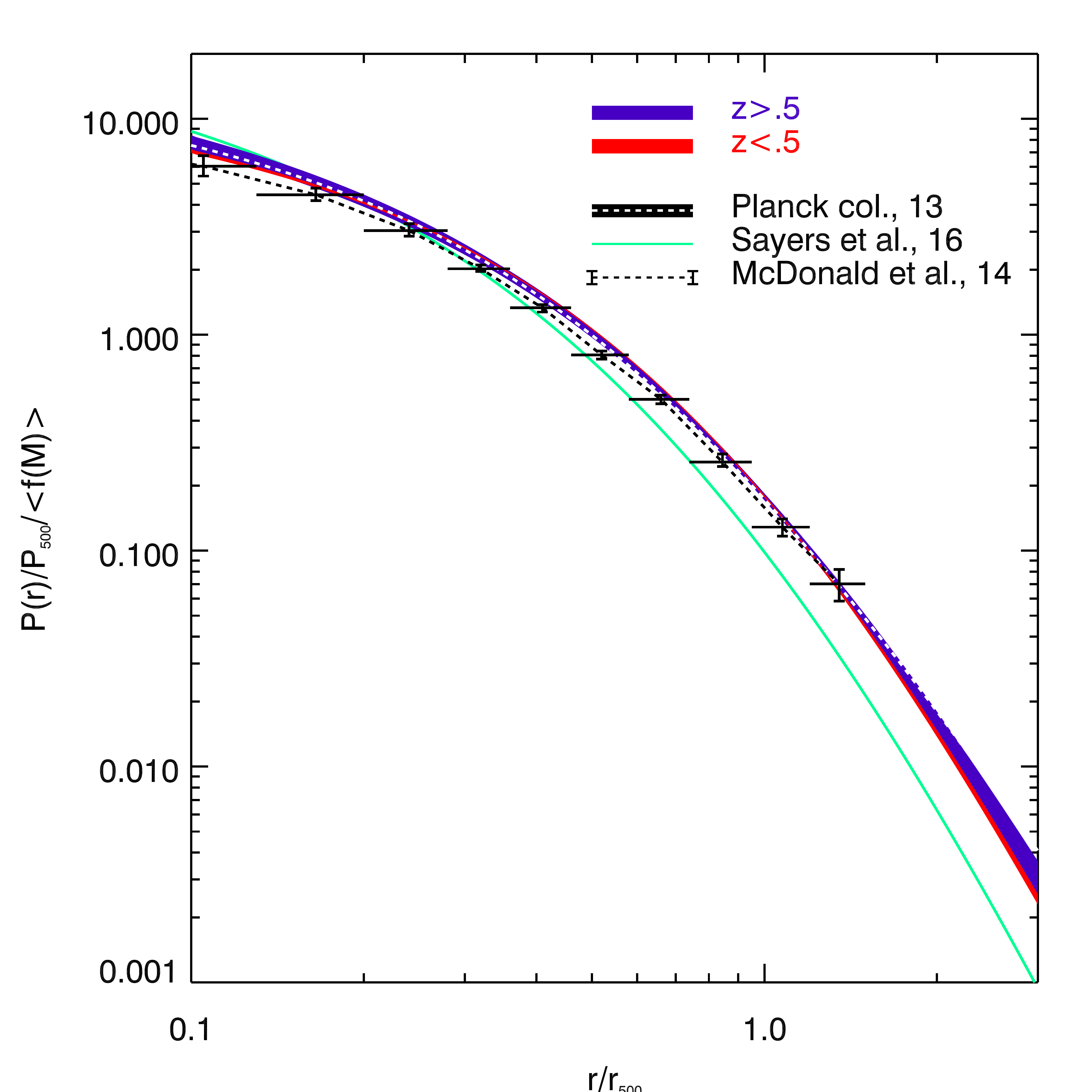}}
  \caption{Mass-scaled pressure profiles in the nearby and distant cluster samples compared with averaged profiles in different millimetric and X-ray analyses.\label{fig:massscaled_pressure_otherworks}}
   \end{center}
\end{figure}

\subsubsection*{Evolution with redshift}

To investigate the redshift evolution of the average profile shape, stacked pressure profiles in both our nearby and distant cluster subsamples are superimposed with one another in Fig. \ref{fig:massscaled_pressure_evol}. Both pressure profiles are compared with the analytical form proposed by A10 that combines X-ray observations of a representative sample of the local cluster population inside $r \sim r_{500}$ with averaged predictions of hydro $N$-Body simulations (\citealt{Borgani_04}, N07, \citealt{Piffaretti_08}) in the cluster outskirts ($r \ge r_{500}$). To take into account a small mass dependence of the profile amplitude that is suggested by the X-ray observations of A10, the stacked profile of each subsample has been renormalized for the arithmetical mean of its member mass functions, $<f(\mathrm{M}_{500})>$, where $f(\mathrm{M}_{500})$ is defined by Eq. (8) of A10:

\begin{equation}
	f(\mathrm{M}_{500}) = \left[ \frac {\mathrm{M}_{500}}{3 \times 10^{14} h_{70}^{-1} \mathrm{M}_{\odot} }\right]^{0.12}. 
	\label{equ:pprof_massfunction}
\end{equation}	

Consistent with results obtained by the \planck~collaboration, the stacked pressure presently derived at low redshift slightly exceeds the average theoretical predictions of A10 beyond $r_{500}$. Moreover, pressure profiles of the nearby and distant subsamples are fully consistent with one another, and a significant excess with respect to A10 is also detected in the outskirts of distant clusters. For cluster masses and distances considered in the present work, we conclude that there is no evolution of the stacked profile shapes or of the individual profile dispersions.

\begin{figure*}[!ht]
   \begin{center}
    \resizebox{\hsize}{!}{\includegraphics{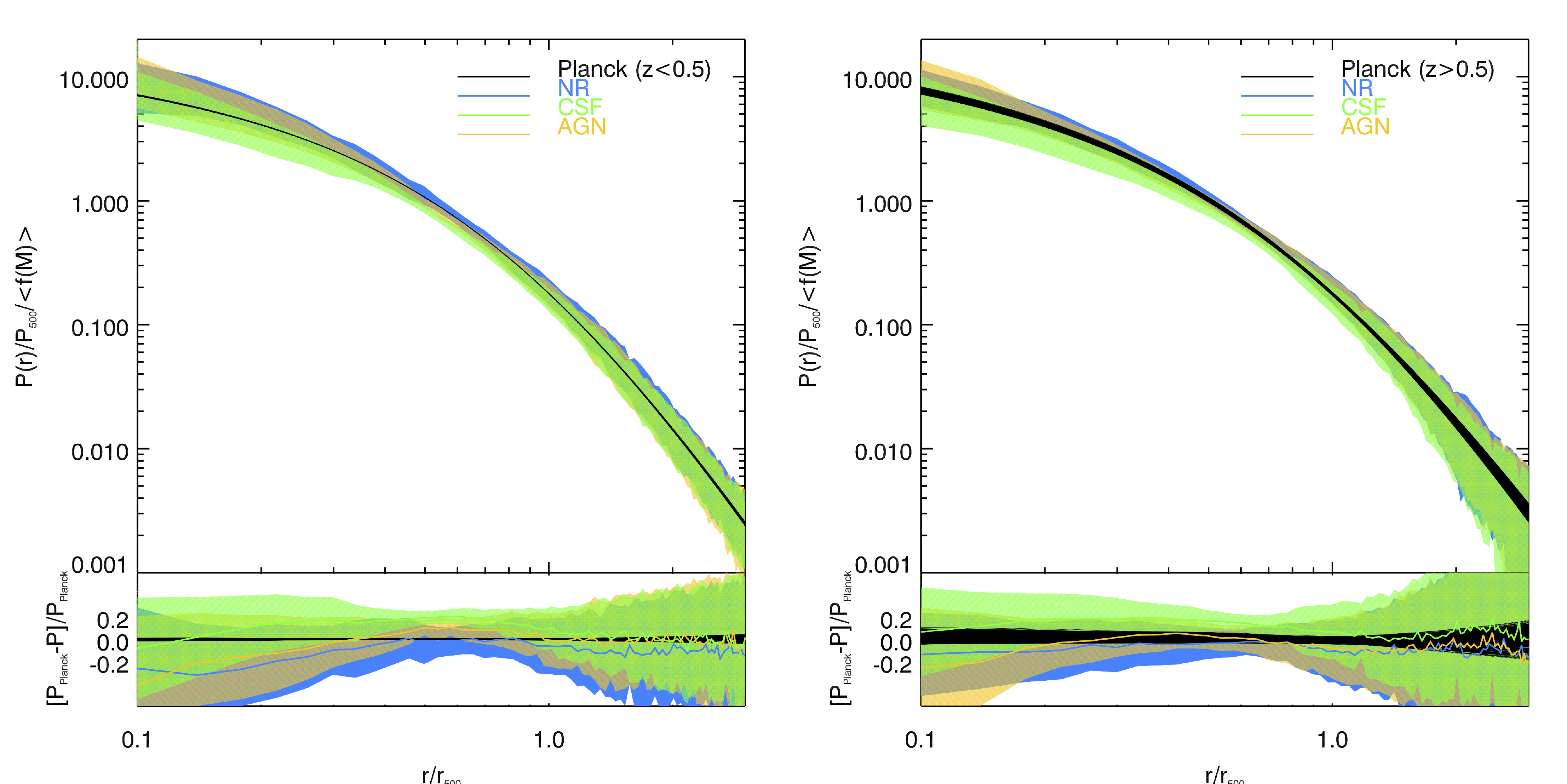}}
  \caption{\textit{Top panels:} Mass scaled pressure profiles in the nearby (left) and distant (right) cluster samples compared with averaged profiles of 24 clusters located at z=0.0 and z=0.5 in the hydrodynamic simulation of \citet{Planelles_17}. Black areas delimit a confidence band over each fitted profile. For each simulation set, continuous curves indicate the azimuthal mean of all profiles, whilst colored envelopes delimit the dispersion of individual profiles relative to their azimuthal mean. \textit{Bottom panels:} Relative difference between simulated profiles and the best profile estimated from \planck~ data. \label{fig:massscaled_pressure_planelles}}
   \end{center}
\end{figure*}

\subsubsection*{Comparison with analogous observations}

Average pressure profiles in our cluster samples are compared to estimates from other observations of moderately distant cluster samples in Fig. \ref{fig:massscaled_pressure_otherworks}. In the millimetric band, \cite{Sayers_16} combined Planck thermal SZ maps derived from the MILCA algorithm with Caltech Submillimeter Observatory observations performed at an arcminute angular resolution near 140 GHz. Using \chandra~observations to derive cluster masses, they derived the averaged pressure profile of a sample of 47 clusters with a median mass of $9.5 \times 10^{14} \mathrm{M_{\odot}} $ and a median redshift of 0.40. This profile has been modeled using the analytical form of Equ. (\ref{equ:nagai_07}) proposed by A10, and fitted to the data using two free parameters: $\mathrm{P}_o=9.13\pm0.68$ and $\beta=6.12 \pm 0.16$. The best estimate of this analysis is represented as a green curve on Fig. \ref{fig:massscaled_pressure_otherworks}. In the X-ray band, \cite{McDonald_14}, used stacked Chandra observations to derive the averaged pressure profile of 40 clusters with an average mass of $5.5 \times 10^{14} \mathrm{M_{\odot}}$ and an average redshift of 0.46. With respect to their published values of $\mathrm{P}/\mathrm{P}_{500}$, data points reported in Fig. \ref{fig:massscaled_pressure_otherworks} have been corrected for the baryon fraction implied by the present definition of $\mathrm{P}_{500}$ (see Eq. \ref{equ:P500}). After being renormalized to their averaged mass function ($<f(\mathrm{M}_{500})>$, see Eq. \ref{equ:pprof_massfunction}), pressure profiles of the present work exceed the profile of \cite{Sayers_16} beyond $r_{500}$, yet appear as fully consistent with the X-ray derived profile of \cite{McDonald_14}.
			
\subsubsection*{Comparison with hydrodynamic simulations}

	Average pressure profiles in our cluster samples are compared with expectations of hydrodynamic simulations in Fig. \ref{fig:massscaled_pressure_planelles}. Specifically, pressure profiles in our low- and high-redshift cluster samples have been superimposed onto the average pressure profile of 24 galaxy clusters with mass $\mathrm{M}_{200} > 8.10^{14} \mathrm{M_{\odot}}$, extracted at  $z=0.0$ and $z=0.5$ from the hydrodynamic simulation sets of \citet{Planelles_17}. Because simulated profiles are not expected to hold any mass-dependent amplitude, no mass renormalization is needed for the simulation set. These simulations proceed from an upgraded version of the TreePM-SPH code GADGET-3 \citep{Springel_05}. They include the updated formulation of Smooth Particle Hydrodynamics presented in \citet{Beck_16}, that take advantage from higher order interpolation kernels and derivative operators than earlier SPH simulations, together with new formulations of artificial gas viscosity and thermal conduction. Three simulation sets corresponding to different ICM physics have been investigated: 1) NR, a non-radiative simulation 2) CSF, a simulation including radiative cooling, star formation, supernovae feedback and metal enrichment 3) AGN, a simulation that adds AGN feedback to the CSF physics. Fig.  \ref{fig:massscaled_pressure_planelles} shows us that the relative differences between pressure profiles simulated assuming these three ICM physics are larger in the innermost cluster regions than in the outskirts. In particular, these differences do not exceed $10 \%$ at $r>r_{500}$, and appear much lower than the cluster-to-cluster dispersion of the profiles. Despite this intrinsic dispersion, the azimuthally averaged pressure profiles in the CSF and AGN simulations coincide remarkably with the stacked pressure profiles observed with \planck~in the cluster outskirts.

\section{Discussion and conclusions}

We combined Planck HFI data with \chandra~and \xmm~observations to extract the thermal SZ signal and pressure profiles of galaxy clusters in the \planck~catalogue.  Assuming spherical clusters, an analytical modeling of the Galactic foreground and CMB anisotropies allows us to deconvolve Compton parameter profiles for the HFI PSF at each frequency. This makes it possible to fully exploit the 5 -arcmin angular resolution of HFI channels beyond 217 GHz and constrain the slope of the pressure profiles in the outer cluster regions ($r \sim 2 \times r_{500}$). The innermost shape and normalisation of the pressure profiles are complementarily constrained using X-ray priors on the radially averaged gas density and the spectroscopic temperature profiles. 

Assuming the spherical symmetry of all thermodynamic quantities that characterize the cluster atmospheres, our normalization of spectroscopic temperature profiles could be used to constrain cosmological models and/or cross-calibrate X-ray and SZ temperature measurements in large cluster samples. Restricting the present work to \xmm~observations of nearby clusters ($z<0.5$) yields a normalization of spectroscopic temperature profiles that is fully consistent with the expected value, $\eta_\T \equiv 1$. In line with earlier results presented by the \planck~collaboration, this suggests that \xmm~and \planck~temperature estimates agree remarkably with one another.

Once renormalized to a characteristic pressure, pressure profiles show a dispersion that remains below $8 \%$ inside $r_{500}$, but increases in the cluster outskirts. Consistent with predictions from $N$-body simulations, profiles thus follow a self-similar behavior driven by cluster masses interior to $r_{500}$. The higher dispersion in the outermost regions reflects a more complex baryonic physics, which is likely perturbed by cluster-to-cluster variations in the abundance of accreting materials. Pressure profiles derived from stacked data sets are fully consistent with the average pressure profile derived by the \planck~collaboration for low redshift clusters ($\tilde{z}=.15$), with stacked \chandra~observations of clusters detected at intermediate redshift with the South Pole Telescope \citep[$\langle z \rangle =0.46$,][]{McDonald_14}, and with expectations from hydrodynamic simulations of \cite{Planelles_17}. After dividing our cluster sample into intermediate ($z<0.5, ~\tilde{z}=0.15$) and low ($0.5<z<1, ~\tilde{z}=0.55$) redshift subsamples, we do not find any evolution of the individual profile dispersion or stacked profile shape. The cluster-to-cluster profile dispersion incites us to investigate the hot gas isotropy and thermodynamics in the outskirts of nearby galaxy clusters that are well resolved with \planck~and accessible to the current X-ray telescopes. It also suggests that any evolution of the profile shapes as a function of cluster masses and distances should be investigated within large cluster samples, possibly via stacking of millimetric and/or X-ray data sets.

\acknowledgments

We wish to thank Susana Planelles for providing us with cluster pressure profiles extracted from her hydrodynamic simulation sets, and Michael McDonald for helpful discussion about the \chandra~follow-up of SPT detected clusters. We thank William Forman, Monique Arnaud and the anonymous referee for useful comments that helped us to improve the manuscript. H.B. acknowledge financial support by NASA grant NNX14AC22G, and Chandra X-ray Observatory grant G05-16147A. P.M. acknowledge financial support by ASI grant 2016-24-H.0. Observations presented in this work were obtained with \planck, \xmm~and the \chandra~X-ray Observatory. \planck~and \xmm~are two ESA science missions with instruments and contributions that are directly funded by ESA Member States, NASA, and Canada.

\bibliography{planck_pressure_apj}

\end{document}